\documentclass[%
 english,
 reprint,
superscriptaddress,
showpacs,
 amsmath,amssymb,
 aps,
prb,
]{revtex4-1}

\usepackage[T1]{fontenc}
\usepackage[latin1]{inputenc}
\usepackage{graphicx}
\usepackage{color}

\usepackage{babel}
\makeatother

\begin{document}

\def\Journal#1#2#3#4{{#1}\ {\bf #2}, #3 (#4)}

\def\CPL{Chem.\ Phys.\ Lett.}
\def\NCA{Nuovo Cimento}
\def\NIM{Nucl.\ Instrum.\ Methods}
\def\NIMA{Nucl.\ Instrum.\ Methods A}
\def\NPB{Nucl.\ Phys.\ B}
\def\PLB{Phys.\ Lett.\ B}
\def\PRL{Phys.\ Rev.\ Lett.}
\def\PRB{Phys.\ Rev.\ B}
\def\PRD{Phys.\ Rev.\ D}
\def\ADNDT{At.\ Data Nucl.\ Data Tabl.}
\def\ZPB{Z.\ Phys.\ B}
\def\ZPC{Z.\ Phys.\ C}
\def\SSC{Solid State Commun.}
\def\SSCH{Solid State Chem.}
\def\JES{J.\ Electr. Spectr.\ Relat.\ Phen.}
\def\NAT{Nature}
\def\AC{Acta Cryst.}
\def\ACA{Acta Cryst.\ A}
\def\ACB{Acta Cryst.\ B}
\def\PB{Physica B}
\def\PC{Physica C}
\def\LCM{Less.-Common Met.}
\def\EPL{Europhys.\ Lett.}
\def\JS{J.\ Super.}
\def\JCG{J.\ Cryst.\ Growth}
\def\JAC{J.\ Appl.\ Crystallogr.}
\def\JPSJ{J.\ Phys.\ Soc.\ Jpn.}
\def\JACS{J.\ Am.\ Chem.\ Soc.}
\def\JLTP{J.\ Low Temp.\ Phys.}
\def\JPCM{J.\ Phys.\ Cond.\ Matt.}



\title{Electronic structure of single-crystalline Mg$_x$Al$_{1-x}$B$_2$ probed by x-ray diffraction multipole refinements and polarization-dependent x-ray absorption spectroscopy}

\author{M.\@ Merz}
\email[Corresponding author: ]{michael.merz@kit.edu}
\affiliation{Institut f\"{u}r Festkörperphysik, Karlsruhe Institute of Technology, 76021 Karlsruhe, Germany}
\affiliation{Institut f\"{u}r Kristallographie, Rheinisch-Westf\"{a}lische Technische Hochschule Aachen, 52056 Aachen, Germany}

\author{P.\@ Schweiss}
\affiliation{Institut f\"{u}r Festkörperphysik, Karlsruhe Institute of Technology, 76021 Karlsruhe, Germany}


\author{Th.\@ Wolf}
\affiliation{Institut f\"{u}r Festkörperphysik, Karlsruhe Institute of Technology, 76021 Karlsruhe, Germany}







\author{H.\@ v.\@ L\"{o}hneysen}
\affiliation{Institut f\"{u}r Festkörperphysik, Karlsruhe Institute of Technology, 76021 Karlsruhe, Germany}
\affiliation{Physikalisches Institut, Karlsruhe Institute of Technology, 76131 Karlsruhe, Germany}

\author{S.\@ Schuppler}
\affiliation{Institut f\"{u}r Festkörperphysik, Karlsruhe Institute of Technology, 76021 Karlsruhe, Germany}

\date{\today}

\begin{abstract}
X-ray diffraction multipole refinements of single-crystalline Mg$_x$Al$_{1-x}$B$_2$ and polarization-dependent near-edge x-ray absorption fine structure at the B 1$s$ edge reveal a strongly anisotropic electronic structure. Comparing the data for superconducting compounds ($x= 0.8$, 1.0) with those for the non-superconductor ($x=0$) gives direct evidence for a rearrangement of the hybridizations of the boron $p_z$ bonds and underline the importance of holes in the $\sigma$-bonded covalent $sp^2$ states for the superconducting properties of the diborides. The data indicate that Mg is approximately divalent in MgB$_2$ and suggest predominantly ionic bonds between the Mg ions and the two-dimensional B rings. For AlB$_2$ ($x=0$), on the other hand, about 1.5 electrons per Al atom are transferred to the B sheets while the residual 1.5 electrons remain at the Al site which suggests significant covalent bonding between the Al ions and the B sheets. This finding together with the static electron deformation density points to almost equivalent electron counts on B sheets of  MgB$_2$ and AlB$_2$\@, yet with a completely different electron/hole distribution between the $\sigma$ and $\pi$ bonds. 
\end{abstract}


\maketitle

\section{Introduction}

\begin{figure}[t]
\hspace{-1mm}
\includegraphics[width=0.5\textwidth]{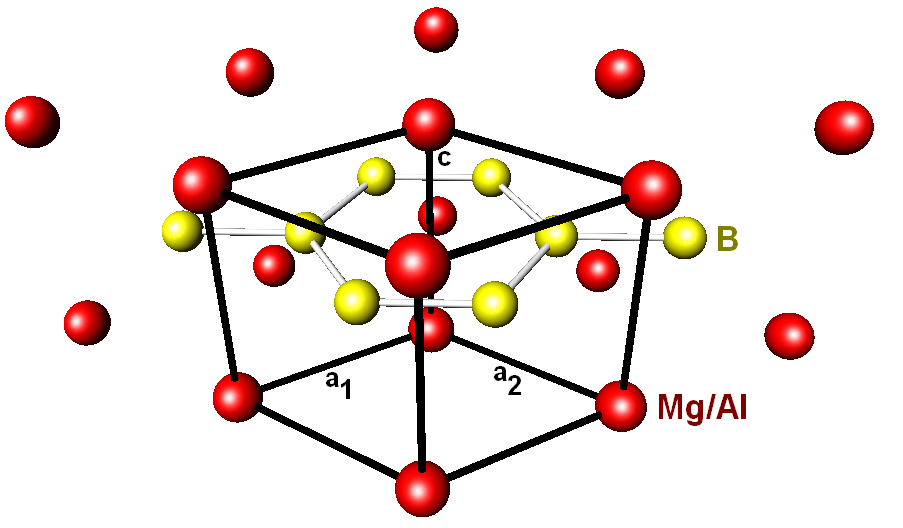}

\caption{\label{fig0}(Color online) Unit cell of Mg$_x$Al$_{1-x}$B$_2$. The red and yellow spheres correspond to the (Mg,Al) and B sites. In its space group $P 6/m m m$ magnesium/aluminum and boron reside on special positions at 0,0,0 and $\frac{1}{3}$,$\frac{2}{3}$,$\frac{1}{2}$\@, respectively. The B rings and the intercalated Mg sheets have a a hexagonal atomic arrangement and are stacked alternatingly along the $c$ axis.}
\end{figure}

Since the discovery of superconductivity in MgB$_2$ with a transition temperature $T_c \approx 39$~K,\cite{akimitsu01} which is high for a binary compound, this material has been intensely studied, and now many properties of MgB$_2$ seem to be well
understood. Compared to the high-$T_c$ cuprates or the recently discovered iron pnictides, the crystal structure is
simple: in its space group $P 6/m m m$ magnesium and boron reside on
special positions at 0,0,0 and $\frac{1}{3}$,$\frac{2}{3}$,$\frac{1}{2}$\@, respectively, and form separate layers which, for B, are graphite-like with a hexagonal atomic arrangement. The B rings and the intercalated Mg sheets are stacked alternatingly along the $c$ axis as illustrated in Fig\@.~\ref{fig0}. This layered structure, and especially the covalent $sp^2$ and two-dimensional character of the B-B $\sigma$ bonds,\cite{an01} 
suggests a strong anisotropy between in-plane and out-of-plane properties.\cite{Guritanu06} Indeed, band-structure calculations predict two degenerate two-dimensional $\sigma$ bands and two three-dimensional $\pi$ bands 
around the Fermi level $E_F$\@.\cite{an01,Kortus01,belashchenko01,suzuki01} Initially it was inferred from experiment and theory that superconductivity in this compound is conventional, being $s$-wave, BCS-type, and phonon-induced.\cite{Budko01,Rubio01,Bohnen01,Renker02} Yet quite early on, deviations were observed, like the presence of two energy gaps of about 7 meV and 2.5 meV related to the $\sigma$ and $\pi$ bands, respectively.\cite{Liu01,Chen01,Szabo01,Laube01} As a result, the superconducting state  of MgB$_2$ is presently understood within a multiband approach of the Eliashberg theory where the pairing is split into $\sigma$ and $\pi$ intraband and $\sigma$-$\pi$ interband contributions.\cite{Kong01,Liu01,Golubov02,Geerk05,delaPena2010} It has been demonstrated that the coupling between the $\sigma$-band electrons and the B-B bond-stretching phonon mode with $E_{2g}$ symmetry at the $\Gamma$ point plays a decisive role for superconductivity in MgB$_2$\@.\cite{delaPena2009,Kong01,Baron2007,Kortus2007,Bohnen01} The electron-phonon coupling for the $\pi$ bands and the interband coupling seem to be weaker although not negligible.\cite{Geerk05}

Much work has concentrated on a direct determination of the
spatial, electronic, and vibrational properties and the superconducting gap structure of MgB$_2$\@. A complementary approach is to investigate systems obtained from the superconducting parent compound by substituting characteristic elements, thereby reducing or even suppressing the superconducting state.\cite{Merz1997} In this respect, the Mg$_x$Al$_{1-x}$B$_2$ system poses an interesting example where superconductivity is suppressed upon replacing formally divalent Mg with formally trivalent Al. Preparation of Mg$_x$Al$_{1-x}$B$_2$ single crystals where Mg is partially or completely replaced by Al is possible for all concentrations. Across the entire doping range $0 \leq x \leq 1$ the crystal structure is preserved while $T_c$\@ is considerably reduced with decreasing $x$ and finally disappears around $x \lesssim 0.6$\@.\cite{delaPena02}
Up to now, several mechanisms like band filling due to substitution of Mg$^{2+}$ with Al$^{3+}$, merging of the superconducting gaps, increased interband scattering, and a decrease in the electron-phonon coupling (Refs.~\onlinecite{Kortus05,Yang03,Gonnelli05,Masui04,delaPena02,delaPena2010,delaPena2009}) have been suggested to explain the very rapidly reduced superconductivity at only modest substitution levels.\cite{Slusky01} So far, however, a consistent picture is elusive and a direct investigation of the electronic structure and the bonding characteristics of the $sp^2$ and $p_z$ states in Mg$_x$Al$_{1-x}$B$_2$ might help in an understanding of the suppression of superconductivity in Mg$_x$Al$_{1-x}$B$_2$ for $x \lesssim 0.6$\@ and of the changes of the electronic structure especially in the less investigated doping range $0 \leq x < 0.6$.

To shed more light on the doping-dependent suppression of the superconducting state and to study the changes in the electronic structure, we have investigated Mg$_x$Al$_{1-x}$B$_2$ single crystals employing x-ray diffraction multipole refinements and polarization-dependent near-edge x-ray absorption fine structure (NEXAFS). X-ray diffraction multipole refinements give direct information about the bonding nature of the relevant electronic states, while NEXAFS tracks the doping-dependent changes in the orbital states and/or in the electronic band structure.  The current data provide experimental evidence that Mg is approximately divalent in Mg$_x$Al$_{1-x}$B$_2$ and suggest a predominantly ionic bonding of the Mg ions and the B layers while Al has a residual valence of $\approx 1.5$ electrons and exhibits significant covalent bonding between the Al atoms and the B sheets. This leads for MgB$_2$ and AlB$_2$ to almost equivalent electron counts on B sites, yet with a completely different electron/hole distribution between the $\sigma$ and $\pi$ bonds. The present data also demonstrate that the doping-dependent filling of $\sigma$-bonded covalent $sp^2$ states goes along with a change of the bonding nature of the $\pi$-bonded $p_z$ orbitals.

\section{Experimental\label{experimental}}

AlB$_2$ and Mg$_{0.8}$Al$_{0.2}$B$_2$ single crystals were grown from
Al or Al/Mg flux by the slow-cooling method. The resulting crystals have hexagonal (001) surfaces with sizes of about 2$\times$2 mm$^2$ for $x=0$ and 0.4$\times$0.5 mm$^2$ for $x=0.8$\@. Hexagonal single-crystalline MgB$_2$ platelets with a diameter of up to 200 $\mu$m are obtained by isothermally annealing $^{11}$B powder in a Mg flux enclosed in an evacuated Mo cylinder.\cite{B11} The MgB$_2$ crystals show a sharp superconducting transition at $T_c=38.8$~K, with a resistive 10 - 90\% width $\Delta T \approx 1.8$~K\@. For Mg$_{0.8}$Al$_{0.2}$B$_2$ the transition is already reduced to around 20~K and considerably
broadened and for AlB$_2$ superconductivity is completely absent.
\begin{table}[b]
\caption{\label{tab:table1} Structural parameters of Mg$_x$Al$_{1-x}$B$_2$ for $x=1.0$\@,
$x=0.8$\@, and $x=0.0$\@. The position of Mg/Al is at $0,0,0$ and that of B at
$\frac{1}{3},\frac{2}{3},\frac{1}{2}$\@; $\alpha = 90^{\circ}$\@, $\gamma = 120^{\circ}$. The
$U_{ii}$ denote the atomic displacement factors ($U_{22}=U_{11}$; $U_{12}= 0.5 \cdot U_{11}$). The
site occupancy factor SOF of B is fixed to 1.}
\begin{ruledtabular}
\begin{tabular}{lcccc}
      &         & $x=1.0$ & $x=0.8$ & $x=0.0$ \\ \hline
      & $wR_2$  (\%)       &  1.85   & 4.44    &  0.97  \\
      & $a$ (\AA)          & 3.085(1)    & 3.071(1) &  3.003(1) \\
      & $c$ (\AA)          & 3.522(1)    & 3.484(1) &  3.254(1) \\
Mg/Al & $U_{11}$ (\AA$^2$) & 0.0051(2)    & 0.0048(1)  &  0.0077(1)\\
      & $U_{33}$ (\AA$^2$) & 0.0057(3)    & 0.0060(1)  &  0.0061(1) \\
      & SOF                & 1.0          & 0.99   &  0.93   \\
      & $\kappa_1$         & 0.89(41)     &     &  1.51(8)  \\
      & $\kappa_2$         & 1.62(41)        &     &  1.30(5) \\
      & $P_{\nu}$         &  0.38(21)     &     &  1.46(10) \\
      & $P_{20}$           & 0.09(3)         &     &  0.43(4) \\
B     & $U_{11}$ (\AA$^2$) & 0.0039(2)    & 0.0039(1)  &  0.0039(1)  \\
      & $U_{33}$ (\AA$^2$) & 0.0061(3)    & 0.0068(2) &  0.0062(1) \\
      & $\kappa_1$         & 0.87(2)      &     &  1.01(2) \\
      & $\kappa_2$         & 0.91(8)      &     &  0.85(4) \\
      & $P_{\nu}$         &  3.82(10)     &     &  3.73(4) \\
      & $P_{20}$          &  -0.07(3)   &     &  0.18(1) \\
      & $P_{33-}$          &  -0.24(4)   &     &  -0.21(3)  \\
Mg/Al--B distance &  $d_1$ (\AA) &  2.505(2)   &  2.486(2)   &  2.378(2)  \\
B--B distance &  $d_2$ (\AA) &  1.781(1)   &  1.773(1)   &  1.734(1) \\
\end{tabular}
\end{ruledtabular}
\end{table}

X-ray diffraction data were collected at room temperature on a Stoe four-circle diffractometer (equipped with a pyrolytic graphite monochromator)
using Mo~$K\alpha_{1,2}$ radiation. All accessible symmetry-equivalent reflections of MgB$_2$,
Mg$_{0.8}$Al$_{0.2}$B$_2$, and AlB$_2$ were measured 
up to a maximum angle $2 \theta$ of $127.5^{\circ}$ ($\sin \Theta / \lambda = 1.26$ {\AA}$^{-1}$).
The data were corrected for Lorentz, polarization, extinction, and absorption effects. 92 and 103 averaged symmetry-independent reflections ($I > 3 \sigma$) have been included for the multipole treatment of MgB$_2$ and AlB$_2$, respectively. The electron density was refined using the Hansen-Coppens formalism~\cite{Jana2000,Hansen78}
\begin{eqnarray}
  \rho_{atom}({\bf r}) &=& \rho_{core}(r) + P_{\nu} \cdot \kappa_1^{3} \cdot \rho_{\rm spher,valence}(\kappa_1
  r)\\ \nonumber
  &+& \sum_{l=1}^{4} \kappa_2^3 \cdot R_l (\kappa_2 r) \cdot \sum_{m=-l}^{l} P_{lm \pm} y_{lm \pm} ({\bf r}/r)
\end{eqnarray}
where $\rho_{core}(r)$ is the contribution from the core electrons;
the second and third term describe the spherical and multipolar
contribution of the valence-electron density, respectively;
$\kappa_1$ and $\kappa_2$ are contraction/expansion parameters of the
valence shell, $R_l$ Slater-type radial functions, and $y_{lm \pm}$
spherical harmonics. The electron populations $P_{\nu}$ and $P_{lm
\pm}$ are determined during the least-squares refinement. According
to the $\bar{6}m2$ symmetry of the B position, $P_{\nu}$\@,
$P_{20}$\@, and $P_{33-}$ values were refined; for the (Mg,Al) site
with $6/mmm$ symmetry $P_{\nu}$ and $P_{20}$ were determined
to correctly describe the valence-electron distribution at this site. 

During the refinement, the high-order reflections with a cutoff of $\sin \Theta / \lambda = 0.9$ {\AA}$^{-1}$ were used to obtain in a first step a good estimate for preliminary anisotropic displacement parameters (ADP's)\@. It should be noted that the atomic positions do not have to be refined since all (Mg,Al) and B atoms reside at special positions in the unit cell. In a second step, the ADP's were fixed and the multipolar refinement of the electron-density parameters 
was performed using all reflections. Finally, the ADP's were relaxed as well and refined together with all multipolar 
parameters. The refinements converged very well and the weighted
reliability factors ($wR_2$) strongly decreased from about 4\% for a conventional refinement to $\approx 1$-2\% for the multipolar treatment. For the multipolar refinements of MgB$_2$ and AlB$_2$, only a marginal residual electron density is found in the
difference Fourier synthesis, which together with the good $wR_2$ values underlines the excellent agreement between the multipole model and the measured data. The multipole parameters obtained in this way were used to calculate 
the static deformation electron density 
which characterizes the redistribution of electrons in relation to a crystal built from spherical atoms\cite{Tsirelson03} and, thus, describes the bonding characteristics of the valence electrons.
Due to the increased disorder expected for the (Mg,Al) position, no multipole modeling was tried for Mg$_{0.8}$Al$_{0.2}$B$_2$.

\begin{figure}[t]
\hspace{-1mm}
\includegraphics[width=0.48\textwidth]{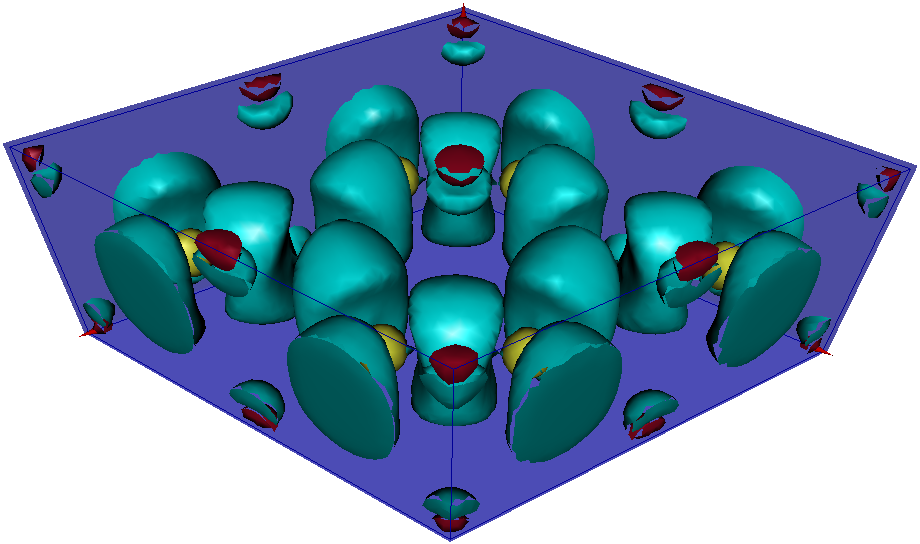}

\caption{\label{fig1}(Color online) Static deformation electron density of MgB$_2$ (turquoise) shown for an isosurface of 0.02 $e$/\AA$^3$. The red and yellow spheres correspond to the Mg and B sites, respectively (see also unit cell in Fig\@.~\ref{fig0}). The electrons predominantly reside on $\sigma$-bonded $sp^2$ hybrids and $\pi$-bonded $p_z$ orbitals, i.~e.\@, between the B sites. The small electron count found at the Mg site might point to a weak hybridization between Mg $3s/3p$ and B $p_z$ states.}
\end{figure}

\begin{figure}[t]
\hspace{-1mm}
\includegraphics[width=0.48\textwidth]{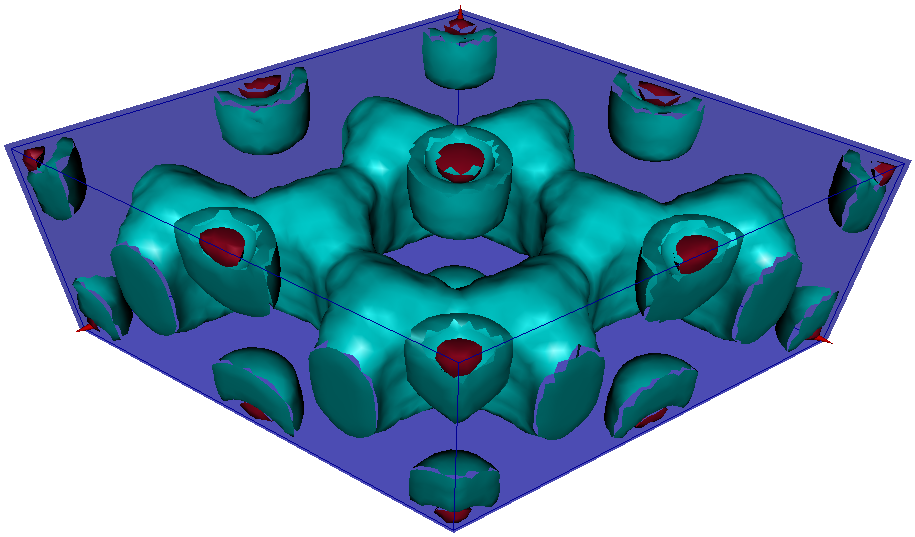}

\caption{\label{fig2}(Color online) Static deformation electron density of AlB$_2$ (turquoise) shown for an isosurface of 0.02 $e$/\AA$^3$. The red spheres correspond to the Al sites (see also unit cell in Fig\@.~\ref{fig0}); the B ions are located below the bulge of the isosurface ring. The $\sigma$-bonded $sp^2$ hybrids are entirely filled with electrons whereas the bulge above and below the B sites together with
the electrons found on Al ions point to a hybridization between Al $3s$/$3p$ and B $p_z$ states.}
\end{figure}
Using linearly polarized light, NEXAFS measurements at the B $1s$ edge were performed for AlB$_2$ at beamline U4B at the National Synchrotron
Light Source (NSLS), Upton, NY, USA, and for Mg$_{0.8}$Al$_{0.2}$B$_2$ and MgB$_2$ at beamline U5 at the National Synchrotron Radiation Research
Center (NSRRC), Hsinchu, Taiwan. Resolution was set to 180 and 100 meV at the
NSLS and the NSRRC, respectively. All data were taken at room
temperature employing bulk-sensitive fluorescence yield (FY), and
were corrected for self-absorption effects using the method outlined in Ref\@.~\onlinecite{Merz1997}\@.
Applying dipole selection rules, the unoccupied part of the B $2p$ final states can be reached from the initial B $1s$ core level. Thus, polarization-dependent NEXAFS measurements on hexagonal Mg$_{1-x}$Al$_{x}$B$_2$ single crystals provide insight into the symmetry of the hole states at $E_F$\@ with B $2p$ character. 
While the in-plane {\bf E}$\perp${\bf c} spectrum is obtained for a normal-incidence alignment, i.~e.\@, for a grazing angle $\theta$\@ of 0$^{\circ}$\@, the out-of-plane {\bf
E}$\|${\bf c} spectrum is determined by measuring in a grazing-incidence setup with a grazing angle of 60$^{\circ}$ and by extrapolating the spectra to $\theta=90^{\circ}$.

\section{Results and Discussion}

The results of the x-ray diffraction refinements are summarized in Table \ref{tab:table1}. The lattice parameters and the ADP's of our single crystals are consistent with established values:\cite{Russell53,Jones76,Nishibori01,Mori02} It is evident that the $a$ and $c$ lattice parameters strongly shrink when going from MgB$_2$ to Mg$_{0.8}$Al$_{0.2}$B$_2$ and AlB$_2$\@. Moreover, our multipole refinement of MgB$_2$ is in agreement 
with published results.\cite{Tsirelson03} In contrast to
Ref.~\onlinecite{Tsirelson03}\@, the Mg site is completely occupied in our MgB$_2$ samples. For the AlB$_2$ sample, on the other hand, the
previously observed slightly reduced 93~\% occupation of the Al position
\cite{Burkhardt04} also shows up in our investigations. According to the
present MgB$_2$ refinement, the valence electron count of Mg is $\approx
0.4(2)$ and that of B $\approx 3.8(1)$\@, indicating that the two valence
electrons of Mg are mostly but not entirely transferred to the B layer. To visualize the valence electron distribution, the static deformation electron density of MgB$_2$ is depicted in Fig.~\ref{fig1}\@. 
It has the shape of a molar and unambiguously resides
between the B sites on $\sigma$-bonded $sp^2$ hybrids and (banana-shaped)
$\pi$-bonded $p_z$ orbitals. Yet both hybrid states seem to be not
\textit{completely} filled (no deformation electron density is found 
around the B sites). Two-dimensional cuts through our three-dimensional plot of the static deformation electron density are absolutely consistent with the deformation density maps given in Ref.~\onlinecite{Tsirelson03}\@. In agreement with Ref.~\onlinecite{Tsirelson03}\@, a small residual number of electrons is found above and below the Mg site (see Fig\@.~\ref{fig1} and Table~\ref{tab:table1}\@). Such a small electron count at the Mg site might point to a weak hybridization between the Mg ion and the B $p_z$ orbitals.

\begin{figure}[t]
\hspace{-2.5mm}
\includegraphics[width=0.45\textwidth]{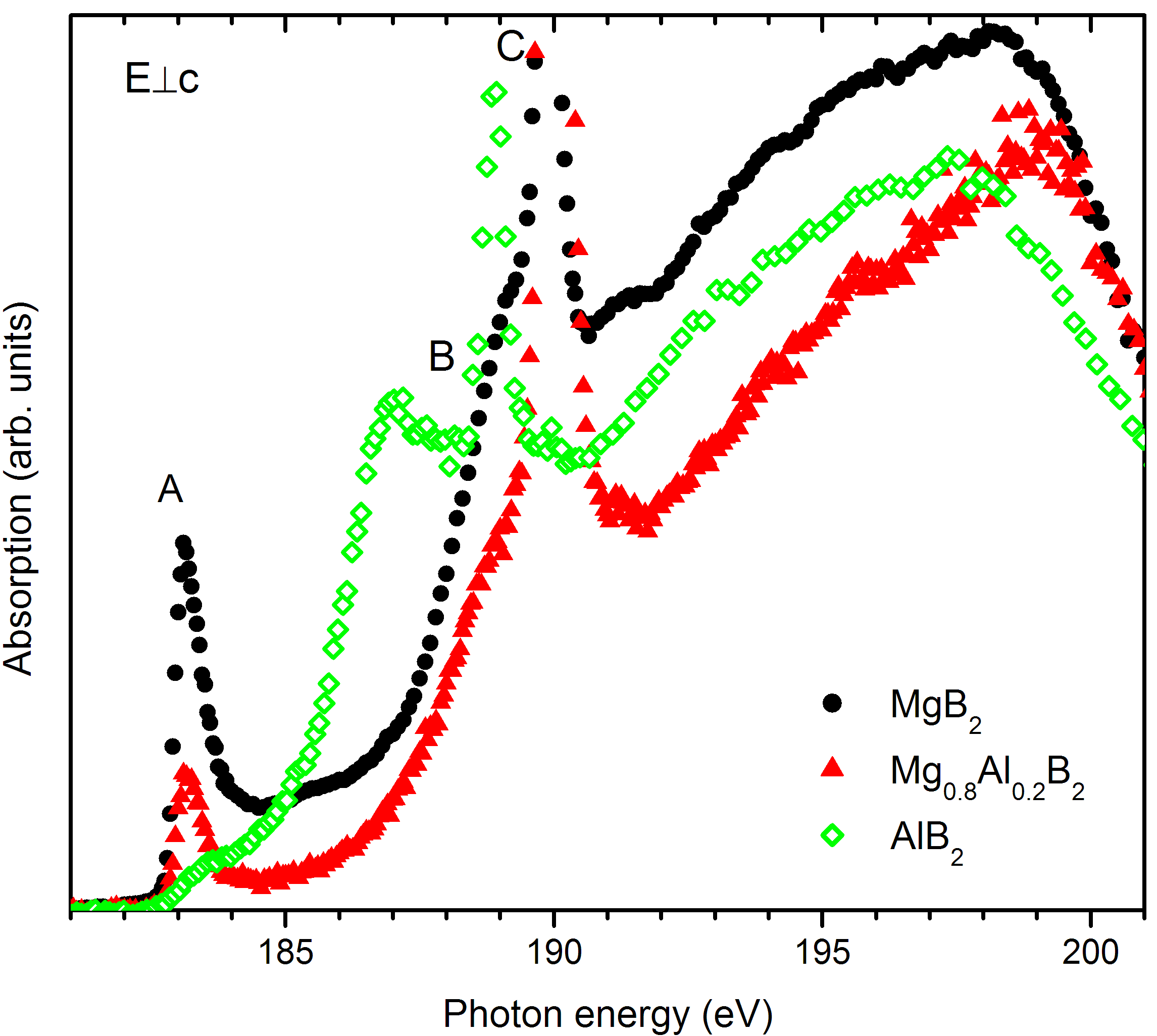}

\caption{\label{fig3}(Color online) Orbital-specific electronic structure of single-crystalline MgB$_2$\@,
 Mg$_{0.8}$Al$_{0.2}$B$_2$\@, and AlB$_2$\@: B 1$s$ NEXAFS with
{\bf E}$\perp${\bf c}, probing unoccupied $\sigma$-bonded B 2$p_{xy}$ and $\pi$-bonded B 2$p_{z}$
states.}
\end{figure}
The situation is, however, quite different for AlB$_2$\@. The
multipole refinement of AlB$_2$ summarized in Table~\ref{tab:table1} clearly shows that not all three 
but only $\approx 1.5(1)$ valence electrons are transferred from the Al site to the B layer while an equal amount of 1.5(1) electrons remains at the Al site. As a consequence this leads to almost equivalent electron counts of $\approx3.8$ for the B layers in MgB$_2$ and of $\approx3.7$ for those in AlB$_2$\@, albeit with a signifantly different electron/hole distribution between the $\sigma$ and $\pi$ bonds. In order to show this, let us compare the static deformation electron density of AlB$_2$ in Fig\@.~\ref{fig2} with that of MgB$_2$ described above and shown in Fig\@.~1\@. For AlB$_2$ the isosurface has the shape of a ring, thereby essentially resembling the hexagonal arrangement of the B ions. Moreover, the comparison of the static deformation electron density of MgB$_2$ and AlB$_2$ suggests that in the case of AlB$_2$ the $\sigma$-bonded $sp^2$ states are completely filled with electrons and shows that the maximum in the electron density of the $p_z$ orbitals does no longer appear \textit{between} the B sites but rather \textit{at} these sites, thereby leading for the isosurface to a bulge above and below the B sites. In addition, a certain electron count is clearly visible around the Al site. Compared to MgB$_2$, the electron density at the (Mg,Al) site is strongly enhanced for AlB$_2$ and, in contrast to MgB$_2$\@, the electron density does not only reside above and below the (Mg,Al) site but rather completely surrounds the Al ion. This finding, together with the 1.5 electrons found on Al sites (see Fig\@.~\ref{fig2} and Table~\ref{tab:table1})\@ and the bulge above and below the B sites, points to strong hybridization between Al $3s$/$3p$ and B $p_z$ states as suggested in Ref.~\onlinecite{Liu2008}\@.
Simultaneously, a non-negligible amount of in-plane holes seems to persist on the $\pi$-bonded $p_z$ orbitals. (Compared to MgB$_2$ a strongly reduced electron density is found between the B sites on banana-shaped
$\pi$-bonded $p_z$ orbitals.) 
Therefore, two effects occur with increasing Al content: Firstly, the $\sigma$-bonded $sp^2$ states become filled with electrons and, secondly, the $p_{z}$ orbitals begin to hybridize strongly with those at the (Mg,Al) site. This second effect changes the bonding nature between the (Mg,Al) site and the B sheets from predominantly ionic for MgB$_2$ to covalent for AlB$_2$\@. These changes are fully consistent with first-principles studies using the virtual crystal approximation where a charge transfer from the B-B $\sigma$ bonds to the $\pi$ bonds in the inter-planar region was observed.\cite{delaPena02,Liu2008} Our data also agree with the detailed supercell calculations in Ref.~\onlinecite{Liu2008} where it was shown that covalent bond clusters composed of Al and the adjacent four B-B bonds emerge upon Al substitution. As a consequence of this increasingly covalent character of the bonds, the (Mg/Al)-B bond length is reduced as well, from 2.505 to 2.486 and 2.378 {\AA} when decreasing the Mg content from $x = 1.0$ to 0.8 and finally 0 (see Table~\ref{tab:table1})\@.

\begin{figure}[t]
\hspace{-2.5mm}
\includegraphics[width=0.45\textwidth]{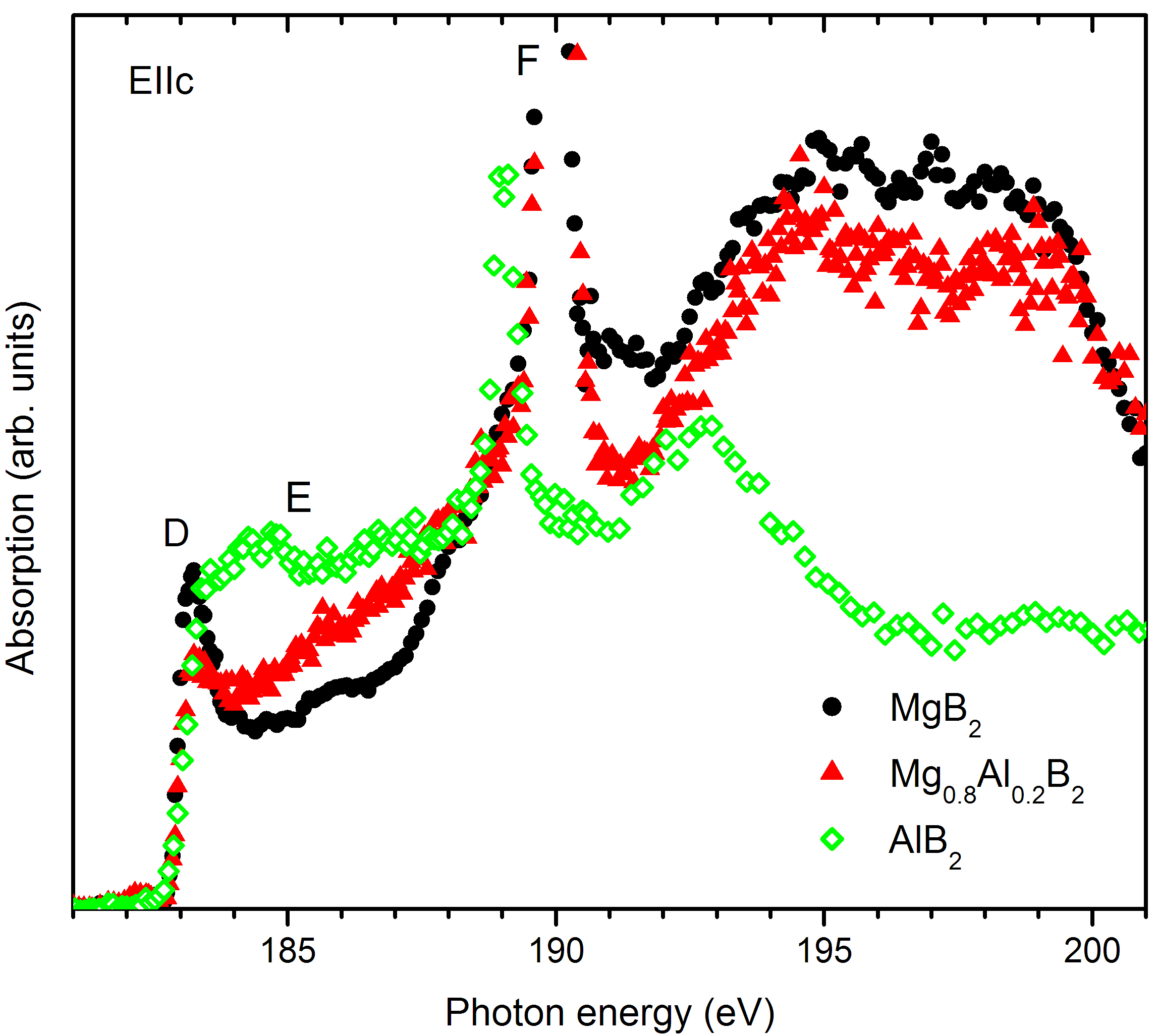}

\caption{\label{fig4}(Color online) Orbital-specific electronic structure of single-crystalline MgB$_2$\@,  Mg$_{0.8}$Al$_{0.2}$B$_2$\@, and AlB$_2$\@: B 1$s$ NEXAFS with {\bf E}$\|${\bf c}, probing unoccupied $\pi$-bonded B~2$p_{z}$ states.} 
\end{figure}
Further support for the successive filling of the $\sigma$ bands and the simultaneous change in the bonding nature of the $\pi$ states comes from the in-plane ({\bf E}$\perp${\bf c}\@) and out-of plane ({\bf E}$\|${\bf c}\@) boron $K$ edge NEXAFS
spectra of MgB$_2$\@,  Mg$_{0.8}$Al$_{0.2}$B$_2$\@, and AlB$_2$\@ displayed between 180 and 201 eV in Figs\@.~\ref{fig3} and \ref{fig4}, respectively. In this energy range the spectra reflect the unoccupied boron $2p$ density of states at $E_F$ and several eV above. For the {\bf E}$\perp${\bf c} spectrum in Fig.~\ref{fig3}, a sharp peak (feature A) appears around $E_F$ for MgB$_2$\@. When going from MgB$_2$\@ to  Mg$_{0.8}$Al$_{0.2}$B$_2$\@ and AlB$_2$\@, a substantial decrease of feature A  is immediately obvious. This reduction of feature A which has already been reported in literature\cite{Yang03,callcott011,callcott012,callcott013,Klie06,Mattila2008}
illustrates that the $\sigma$-bonded boron $sp^2$ states become
gradually occupied upon Al doping and are finally located below
$E_F$ for AlB$_2$. Consistent with the shift of the maximum position
towards the boron sites observed for the static deformation density of
AlB$_2$ in Fig.~\ref{fig2}\@, the residual states found in Fig.~\ref{fig3}\@ for feature A may be attributed to a small amount of in-plane holes residing on $\pi$-bonded boron 2$p_{z}$
orbitals, i.~e.\@, between the boron sites.
Following Ref.~\onlinecite{Klie06}\@, structures B and C of Fig.~\ref{fig3} are ascribed to antibonding $\sigma^*$ states and the so-called
resonance peak, respectively. The latter is either related to oxide impurities or has its origin in resonant elastic scattering \cite{callcott011} and
will not be discussed here. In any case, the antibonding $\sigma^*$
states (feature B) located for MgB$_2$ about 6 eV above $E_F$ (for
$x=0.8$ and 1.0 they only appear as a preceding shoulder of the
resonance peak) are shifted by 1 eV towards
$E_F$\@, but clearly remain unoccupied when going from MgB$_2$ to AlB$_2$\@.

In the out-of-plane ({\bf E}$\|${\bf c}\@) spectra in Fig.~\ref{fig4}\@, a sharp peak (feature D) is observed right at $E_F$\@ for MgB$_2$\@ which can be attributed to $\pi$-bonded $p_z$ orbitals in the B rings. When going from MgB$_2$\@ to Mg$_{0.8}$Al$_{0.2}$B$_2$\@ and AlB$_2$\@, the spectral weight of feature D is significantly reduced while feature E continuously evolves upon increasing Al content, thereby leading for AlB$_2$ to an increased but almost constant density of states in the energy range between $E_F$ and several eV above. Such a constant density of states may point to hybridizations of wide bands. Taking the shown above multipole refinements into account this finding 
reflects the strong hybridization between Al $3s/3p$ and B $p_z$ states. 
Interestingly, feature E already shows up for MgB$_2$\@, yet with significantly reduced spectral weight compared to AlB$_2$\@. As discussed above in the context of the static deformation electron density of MgB$_2$\@, this might point to very weak hybridization between the Mg ion and the B sheets.
Similarly to the in-plane spectra, feature F around 190 eV can be ascribed to
the resonance peak.

Given that Mg$_x$Al$_{1-x}$B$_2$ is a two-band superconductor for which the two-gap characteristic is  preserved over almost the entire superconducting range,\cite{Gonnelli05,Daghero08,Klein06,Szabo07,Gonnelli07} not only the filling of the $\sigma$-bonded $sp^2$ states but also the changes in hybridization between the (Mg,Al) $3s/3p$ and B $p_{z}$ orbitals observed in our experiments might play an important role for the rapid doping-dependent reduction of  $T_c$\@. It was shown (e.~g., in Refs.~\onlinecite{Kortus05,Geerk05,Dahm03}) that superconductivity on the $\sigma$ bands of Mg$_x$Al$_{1-x}$B$_2$ is dominated by $\sigma$-intraband pairing, whereas $\pi$-intraband pairing and  $\sigma$-$\pi$ interband coupling are the prominent ingredients for superconductivity on the $\pi$ bands. Therefore, our data underline that a significant lowering of $T_c$ upon Al doping can be attributed to $\sigma$-band hole filling (and, thus, to reduced $\sigma$-intraband pairing). Simultaneously with the $\sigma$-band filling, the $\sigma$-$\pi$ interband coupling is strongly altered while, at least to first approximation, the $\pi$-intraband pairing is expected to remain unaffected by $\sigma$-band filling. However, the electronic structure together with the $\pi$-intraband pairing will be completely modified by the changes in the hybridization between the (Mg,Al) $3s/3p$ and B $p_{z}$ orbitals observed in our diffraction and NEXAFS results. It might, therefore, be expected that the doping-dependent changes observed in the bonding nature between the (Mg,Al) and the B sites are also an important contribution to the particularly rapid decrease of $T_c$ upon Al doping. The modified electronic structure resulting from the changed bonding nature might also explain the absence of the hole-like $\pi$ sheet for Al-doped samples in de Haas-van Alphen investigations.\cite{Carrington2005} Moreover, such changes in the bonding nature are not expected for the replacement of B with C since the local point symmetry in the $\pi$ and $\sigma$ orbitals is only affected in the case of Mg substitution by Al.\cite{Erwin2003} Indeed the strong reduction of the lattice parameters and of the (Mg,Al)-B bond distance observed for Mg substitution by Al (see Table \ref{tab:table1}) is completely absent for B replacement with C.\cite{Paranthaman2001} Consequently, photoemission studies on Mg(B$_{1-x}$C$_x$)$_2$ indicate that the $\sigma$ gap is proportional to $T_c$ while the $\pi$ gap shows negligible change with increasing B content.\cite{Tsuda2005} This finding for Mg(B$_{1-x}$C$_x$)$_2$ is in contrast to the Mg$_x$Al$_{1-x}$B$_2$ system where, according to our data, it is expected that the $\pi$ gap is correlated with the changes in the covalent bonding nature between (Mg/Al) and B.

\section{Summary and Conclusions}
\label{summary}

To elucidate the doping-dependent suppression of superconductivity in Mg$_x$Al$_{1-x}$B$_2$\@, we have performed multipole refinements of x-ray diffraction data and polarization-dependent near-edge x-ray absorption fine structure on single-crystalline specimens with $x= 1.0,$ 0.8, and 0\@.
Joined together, our results  draw the following picture for the electronic structure and the suppression of superconductivity in Mg$_x$Al$_{1-x}$B$_2$\@: By and large, the Mg ions in MgB$_2$ are  approximately divalent. Despite the residual $\approx 0.4(2)$ electrons remaining at the Mg site, the valence electrons do predominantly reside in
the B layers. Neither the $\sigma$-bonded B $sp^2$ nor the
$\pi$-bonded B $p_{z}$ orbitals are entirely filled with
electrons. This explains the distinctly two-dimensional and covalent character between the B sites with its metallic hole-type conductivity, as well as the appearance of two superconducting energy gaps related to the corresponding
$\sigma$ and $\pi$ bands. 
No indication is found for a three-dimensional, delocalized valence-electron density originating from B~$p_{z}$ bonds; neither the multipole model nor the NEXAFS spectra point to a significant electron
density outside the B layer -- even more so as the distance of about
3.52~{\AA} between
neighboring B layers is quite large which effectively reduces $p_z$ orbital overlap. If anything, the states responsible for three-dimensional metallicity in MgB$_2$ most probably result from 
hybridization between Mg $3s/3p$ and B $p_z$ states. Such hybrids also provide a rationale for the electron density remaining at the Mg site. 

Upon increasing Al content, the $\sigma$-bonded B $sp^2$
states become successively filled. Yet due to the residual valence electron count of $\approx 1.5$ at the Al site as observed in the multipole refinement, the number of electrons in the B layers is only slightly changed from $\approx 3.8$ for MgB$_2$ to $\approx 3.7$ for AlB$_2$. This finding implies that the
$\sigma$-bonded B~$sp^2$ states are filled upon replacing Mg by Al at the expense of an
increasing hole count of the $\pi$-bonded B~$p_{z}$ states as
reflected in the static deformation electron density. Concurrently, the hybridization between (Mg,Al) $3s/3p$ orbitals and B $p_z$ states proceeds with increasing Al content which strongly enhances the covalent bonding
nature between (Mg,Al) and B sites. In contrast to MgB$_2$ where
conductivity is dominated by the in-plane $\sigma$-bonded B $sp^2$
states, in AlB$_2$ the more three-dimensional covalent Al $3s/3p$--B $p_z$ bond determines the electronic properties.

From the present study it is evident that the filling of the
$\sigma$-bonded B~$sp^2$ states has a strong impact on the rapidly reduced superconducting transition temperature in Mg$_x$Al$_{1-x}$B$_2$\@. 
Moreover, it might be expected that the observed changes in the bonding nature between the (Mg,Al) site and the B $p_{z}$ orbitals lead to a particularly rapid decrease of the transition temperature by means of reduced $\pi$-intraband pairing and $\sigma$-$\pi$ interband scattering.
Further detailed experimental and theoretical investigations, however, will be needed to show if the observed changes in the hybridization between (Mg,Al) $3s/3p$ orbitals and B $p_z$ states (and the modified electronic structure of the $\pi$ states caused by these changes) have a decisive role on the superconducting properties of Mg$_x$Al$_{1-x}$B$_2$\@.


\begin{acknowledgments}
We are indebted to G. Roth, G. Heger, and V. Kaiser for fruitful discussions about the multipole refinements. We greatly appreciate generous and excellent experimental help by and helpful discussions with E. Pellegrin, C. T. Chen, S.-C. Chung, S. L. Hulbert, H.-J. Lin, G. Nintzel, S. Tokumitsu, T. Mizokawa, D. A. Arena, J. Dvorak, Y. U. Idzerda, D.-J. Huang, and C.-F. Chang. For stimulating discussions we are grateful to
R. Heid, K.-P. Bohnen, and H. Winter. 
Research was carried out in part at the NSLS, Brookhaven National Laboratory, which is supported by the U. S. Department of Energy, Division of Material Sciences and Division of Chemical Sciences, under contract number DE-AC02-98CH10886.
\end{acknowledgments}
\bibliographystyle{apsrev}
\bibliography{MGB2}

\end{document}